\documentclass[3p,times]{elsarticle}

\usepackage{ecrc}

\usepackage{epstopdf}

\volume{00}

\firstpage{1}

\journalname{Nuclear Physics A}

\runauth{Toru Kojo, Nan Su}

\jid{nupha}

\jnltitlelogo{Nuclear Physics A}

\usepackage{graphicx}
\usepackage{amsmath,amssymb}

\newcommand{\tr}{{\rm tr}}

\newcommand{\lqcd}{\Lambda_{\rm QCD}}

\newcommand{\la}{\langle}
\newcommand{\ra}{\rangle}

\newcommand{\rmd}{\mathrm{d}}

\newcommand{\rme}{\mathrm{e}}

\begin{document}

\begin{frontmatter}



\title{The quark mass gap in strong magnetic fields}

\author[label1]{Toru Kojo}
\author[label2]{Nan Su}

\address[label1]{Department of Physics, University of Illinois, 1110
  W. Green Street, Urbana, Illinois 61801}
\address[label2]{Faculty of Physics, University of Bielefeld, 
D-33615 Bielefeld, Germany}

\begin{abstract}
Quarks in strong magnetic fields 
($|eB| \gg \lqcd^2 \sim 0.04\, {\rm GeV}^2$) 
acquire enhanced infrared phase space
proportional to $|eB|$. 
Accordingly they provide larger chiral condensates 
and stronger backreactions to the gluon dynamics.
Confronting theories with lattice data at various values of $|eB|$, 
one can test theoretical ideas as well as validity of various approximations,
domain of applicability of the effective models, and so on.
The particularly interesting findings on the lattice
are {\it inverse magnetic catalysis} 
and {\it linear growth
of the chiral condensate as a function of $|eB|$},
which pose theoretical challenges. 
In this talk we propose a scenario to explain
both phenomena, claiming that the quark mass gap
should stay at around $\sim \lqcd$, instead of $\sim |eB|^{1/2}$
which has been supposed from dimensional arguments
and/or effective model calculations.
The contrast between infrared and ultraviolet behaviors 
of the interaction is a key ingredient to obtain the
mass gap of $\sim \lqcd$.
\end{abstract}


\end{frontmatter}



\section{Motivations and theoretical problems}
\label{intro}

When uniform magnetic fields are applied to 
a certain direction (we choose the $z$-direction),
quarks wrap around the magnetic field.
Then the orbital motion of quarks in the transverse direction
is quantized, leading to the discretized Landau levels (LLs).
The splitting between each level is $\sim |eB|$.
At tree level, the energy for a quark in the $n$-th Landau level
is given by $(n=0,1,\cdots)$ \cite{Shovkovy:2012zn}
\begin{equation}
E_n(p_z) =\sqrt{\, p_z^2 + 2n|eB| + m^2 \,}\,,
~~~~~~~(m:{\rm quark~mass})
\end{equation}
where the term $2n|eB|$ (we call it transverse energy)
comes from the energies of 
the orbital motion and Zeeman effects.
Each Landau level has a degeneracy factor $\sim |eB|$
with which total number of states remains the same for all $B$. 

What makes QCD in magnetic fields particularly interesting 
is the lowest LL (LLL).
Two aspects are essential in our arguments:
(i) Quarks in the LLL have zero transverse energy
and are indenpendent of $B$ at tree level.
They acquire the $B$-dependence only through interactions.
(ii) For larger $B$, the zero transverse energy and 
Landau degeneracy together allow more quarks
to stay at low energy.
This enhances nonperturbative effects associated with quarks.

On the lattice we can study QCD at various values of $B$,
controlling the size of the infrared phase space for quarks
in the LLL.
Such system can be regarded as a laboratory which is designed 
for the studies of the entanglement between quark and gluon dynamics,
because gluons are affected by $B$ 
only through the couplings to quarks.
Such information would be helpful to understand
the cold, dense QCD matter in which the largest uncertainties come from
treatments of gluons \cite{Rischke:2000cn}.
These considerations greatly motivate the authors to study QCD at 
strong magnetic fields, $|eB| \gg \lqcd^2$.

The lattice results at finite $B$ pose very interesting
theoretical problems 
which are qualitative rather than quantitative.
In this talk we especially focus on two problems as the representatives:
(i) {\it Inverse magnetic catalysis} \cite{Bali:2011qj}.
The (2+1) flavor lattice results for physical pion masses show that 
the critical temperatures ($T_c$) for the chiral restoration
and deconfinement decrease as $B$ increases,
by $10-20\%$ at $|eB| \simeq 1\, {\rm GeV}^2$.
Studies based on the chiral effective models
and the QED-like treatments instead lead to 
increasing critical temperatures;
(ii) {\it The B-dependence of chiral condensates} \cite{Bali:2012zg}.
The effective models or chiral perturbation theories
explain the data well at $|eB|\ll \lqcd^2$, 
but beyond $|eB|\sim 0.1-0.3\, {\rm GeV}^2$,
their predictions start to deviate from the lattice results 
in which the chiral condensate grows like 
$\sim |eB| \lqcd$. 

Below we shall argue how to identify the origin of the problems,
and then give possible resolutions for it.
A key issue will be on the estimate of the dynamically
generated quark mass gap, $M$.
While the estimate based on the dimensional ground
or effective model calculations gives the mass gap of $\sim
|eB|^{1/2}$ \cite{Shovkovy:2012zn},
we instead claim that the mass gap should be nearly $B$-independent
and $\sim \lqcd$, at least within the range of $B$ studied on the lattice.
Then we will outline how to get the mass gap of $\sim \lqcd$
by analyzing the structure of the Schwinger-Dyson equation
at finite $B$ \cite{Kojo:2012js}.

\section{The problems in terms of the quark mass gap}

We are going to phrase the problems in terms of the mass gap.
First we discuss the $B$-dependence of the chiral condensate.
In order to include the $B$ effects to all orders 
for strong $B$ fields, we use the Ritus bases for quarks in the LLs.
With the Ritus bases, we get a formula 
$(p_L =(p_0, p_z), \int_{p_L} \equiv \int \frac{\rmd^2 p_L}{(2\pi)^2}
)$ \cite{Shovkovy:2012zn},
\begin{equation}
\la \bar{\psi} \psi \ra_{ {\rm 4D} }
= \frac{\, |eB| \,}{2\pi}\, \la \bar{\psi} \psi \ra_{ {\rm 2D} }\,,
~~~~~~~~~~
 \la \bar{\psi} \psi \ra_{ {\rm 2D} } 
\equiv - \int_{p_L}
\tr\left[\, S_{ {\rm LLL} }^{ {\rm 2D} } (p_L) 
+ \sum_{n=1} S_{ {\rm nLL} }^{ {\rm 2D} } (p_L)  \, \right]\,.
\end{equation}
The $\la \bar{\psi} \psi \ra_{ {\rm 4D} }$
is the four-dimensional chiral condensate which is proportional to
the Landau degeneracy factor $|eB|$.
After the Landau quantization of the transverse motion, 
quarks in each LL only depend on $p_L$
so that the propagator may be regarded as a two-dimensional one.
We call the chiral condensate made of these two-dimensional propagators
$\la \bar{\psi} \psi \ra_{ {\rm 2D} }$ for the bookkeeping purpose.
The formula purely relies on the bases
and remains valid even after interactions are included.

The propagators of higher LLs
contain inverse powers of $|eB|$,
so the leading $B$-dependence  
of $\la \bar{\psi} \psi \ra_{ {\rm 2D} }$ 
should be dominantly determined by the LLL.
As already mentioned, the LLL acquires the $B$-dependence
only through the interactions and therefore also for the case 
for the resulting dynamical mass gap.
If the mass gap were $\sim |eB|^{1/2}$ as found in effective models,
we would obtain $\la \bar{\psi} \psi \ra_{ {\rm 2D} } \sim |eB|^{1/2}$
(modulo logarithmic $B$-dependence),
which in turn gives $\la \bar{\psi} \psi \ra_{ {\rm 4D} }\sim
|eB|^{3/2}$, faster growth than the linear rising behavior 
found on the lattice.
On the other hand, if we assume the mass gap to be $\sim \lqcd$,
we can get the desired behavior, 
$\la \bar{\psi} \psi \ra_{ {\rm 4D} }\sim |eB| \lqcd$.

The size of the mass gap should also have a big impact
on critical temperatures.
Suppose the mass gap to be $\sim |eB|^{1/2}$ as suggested
by typical effective model calculations.
If quark excitations were such energetic,
the Boltzmann factor would be $\sim \rme^{- |eB|^{1/2}/T }$
for low-lying excitations, 
so that thermal quark fluctuations would not be activated 
until the termperature reaches $\sim |eB|^{1/2} (\gg \lqcd)$.
Thus with this estimate of the mass gap,
the critical temperatures grow like $\sim |eB|^{1/2}$
as $B$ increases.
This increasing behavior is exactly the opposite to the lattice results.
Meanwhile, if we postulate the mass gap to be $\sim \lqcd$,
the Boltzmann factor stays at around $\sim \rme^{- \lqcd/T }$,
with which we can expect the critical temperatures of $\sim \lqcd$
as in the $B=0$ case.
 
Provided that the Boltzmann factor remains the similar 
size as $B$ increases,
the reduction of $T_c$ would not be so surprising.
Having such almost $B$-independent Boltzmann factor,
the major $B$-dependence appears in the number of
possible thermal excitations with the energies below $T$.
At finite $B$, the Landau degeneracy largely enhances
the IR phase space so that
the thermcal fluctuations at finite $B$ 
may be bigger than those at $B=0$.
In our scenario,
it is these enhanced thermal quark fluctuations
that explain the reduction of $T_c$ at finite $B$.

Besides the aforementioned problems,
the estimate of $M\sim |eB|^{1/2}$
would cause troubles in explaining
$10\%-30\%$ changes in the gluonic quantities 
(gluon condensates \cite{Bali:2013esa}, 
string tensions \cite{Bonati:2014ksa}, ...)
at $|eB|\sim 1\,{\rm GeV}^2$ found on the lattice.
Quarks with $M\sim |eB|^{1/2}$ are 
perhaps too massive to affect the gluon sector.

With all these considerations, we expect 
$M\sim \lqcd$ instead of $M\sim |eB|^{1/2}$.

\section{How to get the mass gap of $O(\lqcd)$}

Having stated why the mass gap should be $O(\lqcd)$,
we now consider how to get it.
To compute the mass gap,
we have to solve the Schwinger-Dyson equation.
We include only the LLL because it is the dominant source
for the dynamical generation of the mass gap.
The contributions from the higher LLs
can be safely omitted for 
$|eB| > 0.3\, {\rm GeV}^2$ \cite{Kojo:2013uua}.

The structure of the self-consistent equation for the LLL is given by
\begin{equation}
  M(p_L) \sim \int_{q_L} S_{ {\rm LLL} }^{ {\rm 2D} } (p_L-q_L;M) 
\int_{q_\perp} \rme^{-q_\perp^2/2|eB|}\, D_{ {\rm NP} } (q_L,q_\perp) \,,
\end{equation}
where $S_{ {\rm LLL} }^{ {\rm 2D} } $ is the two-dimensional 
quark propagator for the LLL. 
$D_{ {\rm NP} } (q_L,q_\perp)$ represents interactions
which convolute the nonperturbative gluon propagator and vertex.
Here we have included possible dressing functions for the vertex
into the definition of $D_{ {\rm NP} } (q_L,q_\perp)$.

Several remarks are in order:
(i) The propagator for the LLL depends on $B$ only through the 
mass function $M$;
(ii) The LLL propagator does not manifestly depend on $q_\perp$,
so the integral equation could be factorized;
(iii) In the quark-gluon vertex,
there appears the form factor, $\rme^{-q_\perp^2/2|eB|}$,
because we are using the Ritus bases for quarks 
while the plane wave bases for gluons.
The coupling between different bases yield matrix elements 
with the $B$-dependence incorporated \cite{Kojo:2013uua}.

Note that the $B$-dependence is introduced
only through the form factor $\rme^{-q_\perp^2/2|eB|}$ 
in the second integrand.
Let us define the two-dimensional force, 
\begin{equation}
D_{ {\rm NP} }^{ {\rm 2D} } (q_L; B)\equiv 
\int_{q_\perp} \rme^{-q_\perp^2/2|eB|}\, D_{ {\rm NP} } (q_L,q_\perp) \,,
\end{equation}
which is the four-dimensional force smeared by the
$q_\perp$-integration.
The problem is now reduced to the examination of the
two-dimensional force, because this is the only place
where the $B$-dependence enters.
Our goal is to explain with what kind of forces
the $B$-dependence becomes very weak, leading to 
the estimate $M \sim \lqcd$.

It is instructive to compare several forces.
Let us begin with the contact interaction
for which its strength remains constant from the IR to the UV regions.
In this case the two-dimensional force and the resulting
mass gap become
\begin{equation}
D_{ {\rm contact} }^{ {\rm 2D} } 
= \int_{q_\perp} \rme^{-q_\perp^2/2|eB|}\, {\rm const.}
~\sim~ |eB| ~~~~~ \rightarrow ~~~~~ 
M_{ {\rm contact} } ~\sim~ |eB|^{1/2} \,,
\end{equation}
which are strongly $B$-dependent.
This unwanted parametric behavior is typical for most of effective models.

For the QED-type forces proportional to $1/q^2$
(or the perturbative gluon propagator at fixed coupling),
the $B$-dependence of two-dimensional force
and of the resulting mass gap
is considerably weakened \cite{Gusynin:1995nb},
\begin{equation}
D_{ {\rm QED} }^{ {\rm 2D} } 
= \int_{q_\perp} \rme^{-q_\perp^2/2|eB|}\, 
\frac{\, \alpha_s \,}{\, q_L^2+q_\perp^2\,}
~\sim~  \alpha_s \ln \frac{\, |eB| \,}{q_L^2}
~~~~~ \rightarrow ~~~~~ 
M_{ {\rm QED} } ~\sim~ |eB|^{1/2} \, \rme^{-O(1)/\alpha_s^{1/2} }\,,
\end{equation}
where $\alpha_s \sim \alpha_s(|eB|)$.
The expression for the mass is valid only when  
$\alpha_s$ is small\footnote{In the derivation one keeps only 
the logarithmic terms $\ln \alpha_s^{-1} \gg 1$ while drops
off various $O(1)$ contributions.},
with which the mass gap becomes exponentially small.
Note that the $B$-dependence of the two-dimensional forces is 
only logarithmic, and much weaker than the case of the contact interaction.
To understand this, note that the form factor is activated
when the size of $q_\perp^2$ becomes comparable to $|eB|$,
otherwise it should be close to $1$.
But when $q_\perp^2$ reach $\sim |eB|$, 
the $1/q^2$ force already damps to $\sim 1/|eB|$.
Hence, compared to the case of the contact interaction,
the contributions from the region of $q_\perp^2 \sim |eB|$
become less important.
Although the mass gap from the QED-type forces 
still has the marginal $B$-dependence
and exponentially small,
we are now closer to the desired solution.

Now suppose that $D_{ {\rm NP} }$ in QCD has 
stronger IR enhancement than in the QED case.
Separating the IR contributions from the UV contributions
at some IR scale $\sim \lqcd$,
the two-dimensional force has two distinct contributions,
\begin{equation}
\left( \int_0^{\sim \lqcd^2} + \int_{\sim \lqcd^2}^{\infty} \right)
\rmd q_\perp^2 \, \rme^{- q_\perp^2/2|eB|}\, 
D_{ {\rm NP} } (q_L,q_\perp) 
~~\sim~~ \int_0^{\sim \lqcd^2} \!\rmd q_\perp^2 \,D_{ {\rm NP} } (q_L,q_\perp) 
+ \delta f(p_L, B)
\,.
\end{equation}
Note that for the first integral in the RHS,
we could make replacement, 
$\rme^{- q_\perp^2/2|eB|} \rightarrow 1$, 
for $\lqcd^2/|eB| \ll 1$.
In this way, the $B$-dependence disappears from the first term.
On the other hand, the second term contains marginal $B$-dependence
as seen in the QED case.
But according to our assumption of the IR enhancement,
the second term is negligible compared to the first term
(unless $B$ is extremely 
large\footnote{The UV contribution becomes important when
  $\lqcd \sim |eB|^{1/2} \rme^{-O(1)/\alpha_s^{1/2} } \sim M_{{\rm QED}}$.
This happens when the prefactor $|eB|^{1/2}$
becomes large enough to compensate the exponentially small factor.
We should remember that the expression for $M_{{\rm QED}}$
is valid only for small $\alpha_s$.}
).

It should be noted that the more 
IR enhancement in the $D_{ {\rm NP}}$,
the less $B$-dependence in the resulting two-dimensional force.
If the IR enhancement is large enough,
the two-dimensional force depends upon $B$ only weakly.
Therefore the only relevant dimensionful scale in the 
Schwinger-Dyson equation is $\sim \lqcd$,
so that the resulting mass gap is $M \sim \lqcd$.
This is the desired result.
Tthe recent studies of the Schwinger-Dyson equations 
with nonperturbative forces \cite{Watson:2013ghq}
are in line with the mechanism outlined here.

\section{Conclusions}

Strong magnetic fields
nonperturbatively affect quark dynamics 
by enhancing the IR phase space for quarks.
They provide larger chiral condensates 
and stronger backreactions to the gluon dynamics,
whose qualitative behaviors are different from the predictions
of the effective models for hadron phenomenology.
These discrepancies should not be surprising because
the magnetic field is
much larger than the typical cutoff scale $\sim \lqcd$
in the hadronic descriptions.
In this strong field regime,
proper effective models at finite $B$ can be 
quite different from those at $B=0$.

We have discussed the inverse magnetic catalysis
and the $B$-dependence of the chiral condensate
as the most clear-cut theoretical problems.
The predictions based on effective models and calculations
with perturbative gluons are qualitatively different from
the lattice results.
We attribute the origin of descrepancies to the size of the quark mass gap.
We argued how to get the desired mass gap of $O(\lqcd)$
from the IR enhancement of the QCD forces.

While the regime $|eB|\gg \lqcd^2$ itself is perhaps not directly 
applicable to the QCD phenomenology,
such system can be utilized as an excellent theoretical laboratory.
In particular, QCD at finite $B$ 
has a lot of resemblance to QCD at finite quark density.
We expect that concepts tested and devloped in QCD at finite $B$
can be carried over into the dense QCD matter.

\section*{Acknowledgments}
T.K.is supported in part by NSF Grants PHY09-69790 
and PHY13-05891, and
N.S. is by the Bielefeld Young Researcher's Fund.








\end{document}